\documentstyle[11pt,epsfig]{article}
\columnwidth\textwidth
\newcommand{\lessapprox}{\mbox{$<\approx$}}

\begin{document}
\begin{flushright}
BUTP-9702\\
hep-ph/9702234\\
\end{flushright}

\begin{center}
\Large{\bf  Baryon asymmetry at the weak phase transition
in presence of arbitrary  CP violation.}
\end{center}

\begin{center}
E. Torrente Lujan. \\
Inst. fur Theoretische Physik, Universitat Bern \\
Sidlerstrasse 5, 3012 Bern, Switzerland.\\
e-mail: e.torrente@cern.ch\\
\end{center}

\begin{abstract}
We consider interactions of fermions with the domain wall
bubbles produced during a first order phase transition. A 
new exact solution of the Dirac equations is obtained for 
a wall profile  incorporating a 
 position dependent CP violating phase.
The reflection coefficients are computed,
a resonance effect is uncovered for rapidly varying
phases. This resonance effect happens when
 the energy and mass of the
incident particles are   $E/m=\Delta\theta/2$.
Where $\Delta\theta$ is the phase variation across the wall width.
We calculate the chiral charge flux through the wall surface and the
corresponding baryon asymmetry of the Universe. It agrees in sign
and magnitude with the observed baryonic excess 
$\rho_B/s\approx 10^{-10}$
for a large range of parameters and CP violation.
As a function of $\Delta\theta$, the ratio $\rho_b/s$ reach a maximum 
for $\Delta\theta\approx 2-4\pi$ and $m\approx m_{top}$.


\end{abstract}

\newpage

\section{Introduction}

During the last few  years, a considerable amount of work has been
dedicated to the possibility of generating a sizeable baryon asymmetry
on the electro-weak phase transition (see for example 
\cite{nel1,tou1,aya1,far1}). 
For an excess of baryons to develop in an Universe which initially has zero
baryon number, the already well known 
following conditions, first enunciated by Sakharov, must be
met: 
1) Some interaction of elementary particles must violate baryon number. 
2) C and CP must be violated in order that there is not a perfect equality
between
rates of $\Delta B\not= 0$ processes, since otherwise no asymmetry could evolve
from an initially symmetric state. 
3) A departure from thermal equilibrium must
play an essential role, since otherwise CPT would assure compensation between
processes increasing and decreasing the baryon number. 
Remarkably, the standard model of weak interactions may provide all these
conditions. In particular the third one can be met
if the weak phase transition is at least  weakly first order.

In a first order  phase transition, the conversion from one phase to
the other occurs trough nucleation. This happens when the system is either 
supercooled or superheated. Bubbles of the ''true''
 phase (with an expectation value
of some Higgs field $v\not =0$) expand rapidly absorbing the region of the
''false'' phase ($v=0$). 
At the bubble surface there is a region or ''wall'', in principle of
 microscopic dimensions, which separates the phases.

The speed regime of the expanding bubble walls is poorly understood
in the theory.
After an acceleration period, the final stationary 
 speed
of the expanding bubble walls could be anywhere in the range
 $0.1-0.9 \ c$ \cite{din2}.
Lattice and numerical simulations indicate that $v_{wall}\approx 0.5$
can be reached in circumstances of high latent heat and high surface tension
which could be achieved in the minimal supersymmetric standard model (MSSM)
for  realistic Higgs masses $M_{H}\lessapprox M_W$.
For having even higher wall speed $v_{wall}\approx 0.9$, it is 
necessary some fine tuning (a extremely low value for the friction
parameter, see \cite{kur1}).
At later stages of the bubble evolution an oscillatory regime 
with successive periods of expansion and contraction
is naturally expected
\cite{kur1}.

Particles in the
''false'' (higher temperature) phase are reflected off the advancing bubble 
walls, while most 
particles in the low temperature phase are unable to catch up with the receding
walls, and cannot equilibrate across the phase boundary. Thus one has a 
departure from equilibrium and a baryon asymmetry can be generated.

In the physical conditions of the early Universe the fermions 
moving through
the bubble wall will interact also 
with the particles in
the surrounding plasma,  a full transport problem must 
be considered.
A useful simplifying assumption is to decompose the process into 
two steps,
one describing the production of the CP asymmetry 
on the transmission/reflection coefficients
when the 
quarks/antiquarks
are scattered on the wall, the second describing the transport and the eventual
transformation of the CP asymmetry into a baryon asymmetry via the baryon
number anomaly.

Assuming  that the scattering from the wall is little
 affected by diffusion
corrections, the effects of the surrounding plasma can be
 partially incorporated
by introducing a Higgs field effective potential which 
takes 
into account
finite temperature corrections to the tree-level potential.
The structure and profile
of the wall depends on this effective potential on 
a complicated way. 
Fermions passing through the domain wall acquire a mass which is proportional 
to the  vacuum expectation value (VEV) of the Higgs field,
which is  determined from the equations of motion of the
finite temperature effective action of the bubble.
The problem of computation of transmission coefficients reduces to the 
solution of a Dirac equation with a space dependent mass term. 
Exact solutions has been obtained only 
for two simple cases:
for a wall profile approximated by a step function  \cite{far1,gav1} and for 
an average Higgs field profile of the ''kink ansatz"
type \cite{aya1}(with z, a coordinate perpendicular to the wall):
\begin{equation}
\phi(z)\propto
\left ( 1+\tanh \left (\frac{z}{\delta}\right )\right )
\label{e2001}
\label{e1001}
\end{equation}
where the width of the wall is given essentially by the inverse of the
Higgs mass:
\begin{equation}
\delta=\frac{\surd 2}{M_H}
\label{e2002}
\end{equation}

In  this ansatz
there is not variation of the complex phase of 
the Higgs field through the wall.
It has been examined in \cite{fun2} the effect of introducing a small 
CP-violating imaginary part in Eq.(\ref{e1001})
as a perturbative effect.
Numerical solutions have been obtained for some
more complicated profiles which incorporate 
locally dependent complex phases and which are 
considered to be ''reasonable'' enough although not neccesarily solution of any
equation of motion ( for example in \cite{nel2}).

The object of this work is the computation of transmission coefficients of a fermion in a certain fixed Higgs field background with an arbitrary, possibly high, complex phase. With them we will be
able to estimate the baryon asymmetry of the universe.

Consider a langrangian for the fermion $\Psi$
in the background of a complex scalar with a Yukawa coupling.
 For the scalar to have a nonzero complex vacuum expectation
value, we can assume it is a part of an extended Higgs 
sector such as MSSM or 
a more general two Higgs-doublets model. The fermionic 
part of the lagrangian important to us is
\begin{equation}
L_{\Psi}=\overline{\Psi}_L i\not\partial \Psi_L+
\overline{\Psi}_R i\not\partial \Psi_R
-h\overline{\Psi}_L  \Psi_R \phi
-h^\ast\overline{\Psi}_R  \Psi_L \phi^\ast
\label{e2101}
\end{equation}
with Yukawa coupling constant h. 
$\Psi$ acquires a position dependent mass, which, in 
a appropriate reference frame
is a function only of one coordinate
, $m(z)=h<\phi(z)>$. From  $L_{\Psi}$
the equations of motion are obtained

\begin{eqnarray}
i\not\partial\Psi_L-m^\ast\Psi_R&=&0\nonumber\\
i\not\partial\Psi_R-m\Psi_L&=&0
\label{e2003}
\end{eqnarray}

It  is possible to  absorb a constant 
phase in $m(z)$ with a redefinition
of one of the chiral fields.

One needs only the plane wave solution of the Dirac equation for particles
moving along the normal to the wall profile (considered 
microscopically flat).
For any other incoming direction, the problem can be reduced to the
latter performing a Lorentz boost.
Reordering  the spinorial components,
the Dirac operator 
given by Eqs.(\ref{e2003})
can be factorized into $2\times2$ blocks. 
For solutions with
positive energy E, we can make the ansatz:
\begin{equation}
\Psi=e^{-iEt+i p_t \cdot  x_t}
\pmatrix{ \psi_I(z) \cr \psi_{II}(z)\cr}\equiv
e^{-iEt+i p_t \cdot  x_t}
\pmatrix{\psi_1\cr\psi_3 \cr \psi_4 \cr \psi_2}
\end{equation}

Where $x_t$, is the projection onto the x-y plane.
 $\psi_1$ and $\psi_2$ are eigenspinors of the chirality operator
, $\gamma_5$, for the eigenvalue $+1$ and $\psi_3,\psi_4$ for $-1$.
Going through an intermediate Lorentz transformation
we obtain finally the two equations
\begin{eqnarray}  
(i\partial_z+ Q(z))\psi_I&=&0 \label{e1102}\\
(i\partial_z+ Q^\ast(z))\psi_{II}&=&0 
\label{e1101}
\end{eqnarray}
With 
\begin{equation}
Q(z)=\pmatrix{E_l & -m(z) \cr m(z)^\ast & -E_l};\;
E_l=+\sqrt{E^2-p_t^2}
\end{equation}

We will deal in this work with the particular case given by the function
\begin{equation}
m(z)=\left\{ 
\begin{array}{ll}
m_0 \exp i( -\Delta \theta \lambda z) \exp -\lambda z&
 \mbox{if $z>0$} \\
m_0 & \mbox{if $z\leq 0$}
\end{array}\right .
\label{e1103}
\end{equation}
The constants $\Delta\theta,\lambda$ are real;
 $\lambda >0 $ ;
 $m_0$ can be also   set 
to real without loss of generality.
 This function
represents a
linear phase variation with a global
 difference of $\Delta \theta$ over a distance of 
the order of the wall thickness defined by 
$\delta\equiv 1/\lambda$. In specific models $\Delta\theta$ can be related to 
the quantity of CP violation in the Higgs sector.
The position dependent complex phase can be considered as  an additional source 
of effective CP-violating phenomena.

The profile of the wall 
given by $m(z)$
cannot be derived from any classical field
equations   in presence  of an 
effective Higgs potential with up to quartic terms as it is 
usually considered. It can be trivially seen 
however
that such a wall 
profile corresponds to an energy density of the wall per unit area of the form
\begin{equation}
{\cal E}(\Delta\theta)-{\cal E}(0)= \lambda m_0^2 \Delta\theta^2
\end{equation}
with ${\cal E}(0)$ of the same order of magnitude 
 as the energy density of the 
''kink'' ansatz (\ref{e2001}).
Some indication can be obtained from the previous formula.
As the nucleation temperature $T_n\propto \cal E$
(see Eq.(13) in \cite{abn1}) and according to
Table (1) in \cite{kur1} which relates nucleation rates and wall speed, 
a high value for $\Delta\theta$ would favor generically a 
lower
value for the wall speed. In this work we will see (Fig.(6)(C-D)) that, for small wall speed, a high value for $\Delta\theta$ is equally acceptable as a
lower one in order to obtain a reasonable $\rho_B/s$  ratio.

\section{Solving the Dirac equation.}
It would be possible to solve Eqs.(\ref{e1102}-\ref{e1101}) with the function
given by Eq.(\ref{e1103}), reducing them to a  
Bessel-like differential equation. It is possible and advantageous 
however to use perturbation theory to compute the evolution operator of the
system. The summation to all orders of the perturbation expansion is possible
thanks to the special form of $Q(z)$. 
There are two main advantages in doing so:
the first one is that
the procedure is easily generalizable to any number of dimensions (for 
example to incorporate mixing between generations), the second one 
is that the
evolution operator is computed directly and the reflection coefficients
are easily given in terms of its components. 
A similar technique has been used  already in \cite{emi1,emi2} to compute 
the neutrino 
oscillation probabilities in solar matter.

The evolution operator of the differential system  (\ref{e1102})
is given by the path-ordered integral
\begin{equation}
U(z,z_0)=P\exp i\int_{z_0}^z dz Q(z)
\label{e102}
\end{equation}
In this work we are concerned with an operator $Q$ of the form
\begin{equation}
Q=Q^0+  V(z)
\label{e6009}
\end{equation}
With
\begin{equation}
\begin{array}{cc}
Q^0=\pmatrix{E & 0 \cr 0 & -E}; & V=\pmatrix{0 & -k \exp - \sigma z\cr
k\exp - \sigma^\ast z& 0}
\end{array}
\label{e1110}
\end{equation}

$k$ real, $\sigma$ complex in general. 
 The real part of $\sigma$ is greater than zero. It will be
 set $\Re \sigma=1$ without loss of generality.

 Formally, it is 
possible to solve  Eq.(\ref{e102}) by successive iterations:

\begin{equation}
U(z,z_0)=U^{(0)}(z,z_0)+\sum_{n=1}^{\infty}U^{(n)}(z,z_0) ;\quad
U^{(0)}(z,z_0)=\exp\left (i Q^{0}(z-z_0)\right )
\end{equation}

 $U^{(n)}$ is the well-known integral
\begin{equation}
U^{(n)}=i^n\int_{\Gamma}
dz_n
\dots dz_1 
U^0(z,z_n)V(z_n)
\dots
U^0(z_2,z_1)V(z_1)U^0(z_1,z_0)
\label{e101}
\end{equation}

The domain of integration is defined by 
$$\Gamma\equiv z>z_n>\dots >z_1>z_0.$$

Following the same arguments as in \cite{emi1} one can show that it is
enough to compute
U in the $z\rightarrow\infty$ limit.
The evolution for finite $z$ can be deduced from the expression for this
limit. 
 Trough elementary manipulations of Eq.(\ref{e101}), we get the 
elements of $U^{(n)}$ in a basis of eigenvectors of $Q^0$:
\begin{eqnarray}
\lefteqn{<b\mid U^{(n)}\mid a> =  
i^n\exp(i (Q_b^0 z-Q_a^0 z_0)) \times}\nonumber \\
 & &\times \sum_{k1,\dots,k(n-1)}\int_{\Gamma}
d^n\tau
\exp(iz_n w_{bk1}+\dots+ iz_1 w_{k(n-1)a} ) 
V_{bk1}(z_n)\dots V_{k(n-1)a}(z_1) 
\label{e1107}
\end{eqnarray}
With $w_{k1k2}=Q_{k1}^0-Q_{k2}^0$. $Q^0_k$ one of the two eigenvalues of
$Q^0$.

Due to the  dimensionality of the problem and the special form for
V, the summatory in Eq.(\ref{e1107}) either is zero or reduces to only one term
depending on whether n is odd or even and on the states $a,b$. 
For diagonal terms ($a=b$) the 
 product of $V\dots V$ is always 
zero when n is odd. When n is even there is a single  surviving term.
For non-diagonal terms, the situation is reversed: the
 only one surviving term appears for n odd.
This single 
remaining term is always of the alternating form  
$\dots V_{12}V_{21}V_{12}\dots $. So,
\begin{equation}
<b\mid U^{(n)}\mid a> =  
i^n\exp(i (Q_b^0 z-Q_a^0 z_0))I_{ab}^{(n)}\times  
\left \{\begin{array}{cc}
\left (-\mid k\mid\right )^{n/2}  &  a=b \\
V_{ba} \left (-\mid k\mid\right )^{(n-1)/2} &  a\not=b 
\end{array}
\right .
\label{e6020}
\end{equation}

All the functions appearing in the integral $I_{ab}^{(n)}$
are of exponential type,
the following equality \cite{emi1} can be applied:
\begin{eqnarray}
I_n( w_1,\dots,w_n) &
\equiv &
\int_{z_0}^\infty\dots\int_{z_0}^{x_2} dx_n \dots dx_1\exp
\sum_n w_n x_n \nonumber\\
& =&
   \frac{ (-1)^n \exp (z_0\sum_n w_n)}{w_1 (w_1+w_2)\dots 
(w_1+w_2\dots +w_n)}\nonumber \\
& & \nonumber\\
& & (\hbox{valid if\ }\ \  \Re{\ w_n} < 0,\ \forall\ n\ ) \label{e612}  
\end{eqnarray}

In our case the sums in the denominator of Eq.(\ref{e612}) are of the form

\begin{equation}
\begin{array}{rl}
w_1+w_2+\dots + w_j &= i w_{bk1}-L+ i w_{k1k2}-L^\ast+\dots +i
w_{k(j-1)kj}-L\\
 &= iw_{bkj}+n_1 L+n_2 L^\ast
\end{array}
\end{equation}

$w_{bk(j)}$ can take only the values $\{0,\pm 2E\}$. $L$ is
$\sigma$ or $\sigma^\ast$. One have for the integers $n_1,n_2$: $n_1+n_2=j$
$n_1=n_2$ or $n_1=n_2\pm 1$; they are equal or differ in one unit.

For the diagonal terms, $a=b$, $n$ even  (writing $s=2Ei-\sigma$): 
\begin{eqnarray}
I_{aa}^{(n),even}&=& 
\frac{\exp z_0 (n/2)(s+s^\ast) }
{s^\ast  (s^\ast+s) (s^\ast+s+s^\ast)\dots((n/2) (s+s^\ast))}
\nonumber \\
&=& \frac{\exp z_0 (n/2)(s+s^\ast) }
{\prod_{j=2,even}^n (j/2) (s+s^\ast) \prod_{j=1,odd}^{n-1} 
[s^\ast+(j-1)(s+s^\ast)/2]}
\nonumber \\
&=&\frac{\exp z_0 (n/2)(s+s^\ast) }
{(s+s^\ast)^{n/2} \left(\frac{n}{2}\right )! (s+s^\ast)^{n/2}\prod_{l=1}^{n/2} [
s^\ast/(s+s^\ast)+ (l-1)] }
\nonumber \\
&=&\frac{\exp z_0 (n/2)(s+s^\ast) }
{(s+s^\ast)^n \left(\frac{n}{2}\right )!  \left [s^\ast/(s+s^\ast)\right 
]_{(n/2)}}
\label{e6019}
\end{eqnarray}

For non-diagonal terms, taking  $a=1,b=2$ to simplify the notation:
\begin{eqnarray}
I_{ab}^{(n),odd}&=& 
\frac{(-1)\exp z_0 ((n/2)(s+s^\ast)+s^\ast) }
{s^\ast  (s^\ast+s) (s^\ast+s+s^\ast)\dots((n/2) (s+s^\ast)+s^\ast)}
\nonumber \\
&=& \frac{(-1)\exp z_0 (n/2)(s+s^\ast) }
{\prod_{j=2,even}^{n-1} (j/2) (s+s^\ast) \prod_{j=1,odd}^{n} 
(s^\ast+(j-1)(s+s^\ast)/2)}
\nonumber \\
&=&\frac{(-1)\exp z_0 (n/2)(s+s^\ast) }
{(s+s^\ast)^{(n-1)/2} \left(\frac{n-1}{2}\right )! 
(s+s^\ast)^{(n+1)/2}\prod_{l=1}^{(n+1)/2} [s^\ast/(s+s^\ast)+ (l-1)] }
\nonumber \\
&=&\frac{(-1)\exp z_0 (n/2)(s+s^\ast) }
{(s+s^\ast)^n \left(\frac{n-1}{2}\right )!  \left [s^\ast/(s+s^\ast)\right 
]_{(n+1)/2}}
\label{e6018}
\end{eqnarray}

We have used the Pochammer symbol defined by Eq.~(\ref{a110}).

So, inserting Eqs.(\ref{e6019},\ref{e6018}) in Eq.(\ref{e6020}) and taking 
$s+s^\ast=-2$, $z_0=0$:

\begin{eqnarray}
\lefteqn{<1\mid U\mid 1>=} \nonumber \\
 &= & 
e^{i E z}  \left ( 1+\sum_{n=2,even}^{\infty} 
i^n \left (-k^2\right )^{n/2}   
\frac{1}
{(-2)^n (\frac{n}{2})!  [ -s^\ast/2]_{(n/2)} }\right )
 \nonumber \\
 &= & e^{i E z}
\left ( 1+ \sum_{m=1}^\infty
\left (\frac{ k^2}{4}\right )^m \frac{1}{m! [-s^\ast/2]_{(m)}}\right )
 \nonumber \\
&=& e^{ i E z} 
{}_0 F_1 \left (- \frac{s^\ast}{2}; \frac{k^2}{4}\right )
\end{eqnarray}

\begin{eqnarray}
\lefteqn{<1\mid U\mid 2>=} \nonumber \\
 &= & e^{ i E z}  
k\sum_{n=1,odd}^{\infty} 
i^n \left (-k^2\right )^{(n-1)/2}   
\frac{(-1)}{
(-2)^n (\frac{n-1}{2})! [ -s^\ast/2]_{((n+1)/2)}}
\nonumber \\
&=& e^{i E z}\frac{-2ik}{k^2}   
\sum_{m=1}^{\infty}    
\left (\frac{k^2}{4}\right )^m  
 \frac{1}{\left(m-1\right
)! [-s^\ast/2]_{(m)}} \nonumber \\
&=&e^{i E z} \frac{-ik}
{s^\ast}\
{}_0 F_1\left (1-\frac{s^\ast}{2};
\frac{ k^2}{4}\right )
\end{eqnarray}
and similarly for the other two matrix elements.
\begin{eqnarray}
<2\mid U\mid 2>
&=& e^{ -i E z} 
{}_0 F_1 \left (- \frac{s}{2}; \frac{k^2}{4}\right )
\end{eqnarray}

\begin{eqnarray}
<2\mid U\mid 1>&=& e^{ -i E z} 
 \frac{ik}{s}\ {}_0 F_1\left (1-\frac{s}{2};
\frac{ k^2}{4}\right )
\end{eqnarray}

The matrix U can be written as 
\begin{eqnarray}
U(z\to\infty,z_0)&=&\exp i Q_0 (z-z_0) U_{red}( z_0)\nonumber \\
     & & \nonumber \\   
U_{red}(z_0)&=&\pmatrix{
 F &  G \cr G^\ast & F^\ast \cr }
\label{e2120}
\end{eqnarray}

writing
 $\sigma=\lambda (1+i\Delta\theta)$, $k=m_0/\lambda$, $E=E_l/\lambda$
\begin{eqnarray}
 G&=&\frac{-i m_0/\lambda }
{1+(2E_l/\lambda-\Delta \theta) i} {}_0 F_1\left (\frac{3}{2}+(\frac{E_l}{\lambda}-\frac{\Delta \theta}{2})
 i ; \frac{m_0^2}{4\lambda^2}\right) \nonumber \\
F&=&{}_0 F_1\left (\frac{1}{2}+(\frac{E_l}{\lambda}
-\frac{\Delta \theta}{2}) i; 
\frac{m_0^2}{4\lambda^2}\right  )
\label{e6024}
\end{eqnarray}
In the case $z_0\not= 0$, $F,G$ would include a factor $\exp -2z_0$ in its
argument. 
The special case  $E_l/\lambda=\Delta\theta/2$ will be important later. 
In this case $F,G$ become dependent only on $m_0$ and have the explicit 
simple form:
\begin{eqnarray}
 G&=&-i \frac{m_0}{\lambda} 
{}_0 F_1\left (\frac{3}{2} ; \frac{m_0^2}{4\lambda^2}\right) =-i \sinh\left(
\frac{m_0}{\lambda}\right ) \nonumber \\
F&=&{}_0 F_1\left (\frac{1}{2};\frac{m_0^2}{4\lambda^2}\right  )
=\cosh\left (\frac{m_0}{\lambda}\right )
\label{e6024b}
\label{e1028}
\end{eqnarray}

To obtain $\overline{U}$, evolution operator for the Eq.(\ref{e1101}), one must
make the change $\Delta\theta\to -\Delta\theta$.  The matrix become
\begin{eqnarray} 
\overline{U}(z\to\infty,z_0)&=&\exp i Q_0 (z-z_0)
\overline{U}_{red}( z_0)\nonumber \\ & & \nonumber \\
\overline{U}_{red}(z_0)&=&\pmatrix{ \overline{F} & \overline{G} \cr
\overline{G}^\ast & \overline{F}^\ast \cr } \label{e2122} 
\end{eqnarray} 
with
\begin{eqnarray} 
\overline{G}&=&\frac{-im_0/\lambda}{1+(2E_l/\lambda+\Delta \theta) i} \ 
{}_0 F_1\left (\frac{3}{2}+(\frac{E_l}{\lambda}+\frac{\Delta \theta}{2}) i ;
\frac{m_0^2}{4\lambda^2}\right)\nonumber \\ \overline{F}&=&{}_0 F_1\left
(\frac{1}{2}+(\frac{E_l}{\lambda}+\frac{\Delta \theta}{2}) i; \frac{m_0^2}{4\lambda^2} \right )
\label{e6025} 
\end{eqnarray}

See Appendix A for some new formulas for the absolute values of generalized
hypergeometric functions which can be obtained from the general properties of
$U,\overline{U}$.

Following the same reasoning used in \cite{emi1}, using the general properties
of the evolution operator, the evolution for any finite z is given by the matrix
\begin{eqnarray} U(z,z_0)&\equiv&U_s(z)^{-1}U_{s}(z_0)=U_{red}^{-1} (z) e^{i Q_0
(z-z_0)} U_{red}(z_0) \label{e1125} \end{eqnarray} In this work, we will make
use only of the infinite $z$ limit.

     \section{ The reflection Coefficients.}

     We will follow the same procedure as in \cite{nel1} for defining the
     reflection coefficient.  As it can be seen directly from
     Eqs.(\ref{e2120},\ref{e2122}), in the region $z\to\infty$ we have the
     following asymptotic behavior
  
     \begin{eqnarray}
     \psi_2(\infty),\psi_3(\infty)&\simeq& e^{-iE_l z}\nonumber\\
     \psi_1(\infty),\psi_4(\infty)&\simeq& e^{iE_l z} 
     \end{eqnarray}

     In this region , the eigenspinors
     $\psi_2,\psi_3$ correspond to, vanishing mass, left-moving particles of
     opposite chirality and similarly $\psi_1,\psi_4$ to right moving particles.

     For left-moving particles incident from the symmetric phase two components
      coexist at $z\to\infty$, the incident particle itself and the reflected
      one by the domain wall.  We define asymptotic reflection coefficients
      $R,\overline{R}$ as 
      \begin{equation} 
      \psi_1(\infty)=R \psi_3 (\infty);
      \quad \psi_4(\infty)=\overline{R} \psi_2(\infty) 
       \end{equation}

     The momentum eigenstates for $z<0$ are obtained diagonalizing the constant
     matrix $Q(0)$.  The eigenvectors 
     \begin{equation} 
     \psi_{\pm}=\frac{1}{\lambda}\pmatrix{ m_0 \cr E_l-p_{\pm} } 
     \end{equation} 
     correspond to right moving (+) and left
     moving particles (-) with respective eigenvalues \mbox{$p_\pm=\pm
     p_l=\pm\sqrt{E_l^2- m_0^2}$}.

     Imposing the boundary condition that at $z<0$ only a left-moving particle
      with momentum $p_{-}$ propagates, the wave function at infinite is:
      \begin{equation} \psi(\infty)=U(z\to\infty,0) \psi_{-} \end{equation}

     The reflection coefficient R is given explicitly by the expression
      \begin{equation} 
      R=\frac{U_{11} +t U_{12}}{U_{21}+t U_{22} }=e^{2E_l zi}
      \frac{F+t G}{G^\ast+ t F^\ast}; \quad 
      t=(E_l+p_l)/m_0
      \label{e2129}
      \end{equation} 
      and similarly for $\overline{R}$.  

     We are interested in the asymmetry between the reflection probabilities:
     $A=\mid R\mid^2-\mid \overline{R}\mid^2$.
      In the ultrarelativistic
      limit ($E_l/m_0>>1$), $R\to F^\ast/G$. 
      In the limit $E/ m_0\to 1$ $(t\to 1)$, both reflection coefficients
      $\mid R\mid,\mid \overline R\mid\to 1$ and the asymmetry $A\to 0$.
     It is straightforward to check that in the case we
     had supposed from the beginning a complex constant mass $m_0$, 
    the quantities $R,\overline{R}$
      would remain independent of the phase of $m_0$ taking into account the
     formula (\ref{e2129}).

\begin{figure}[p]
\centering\hspace{0.8cm}
\epsfig{file=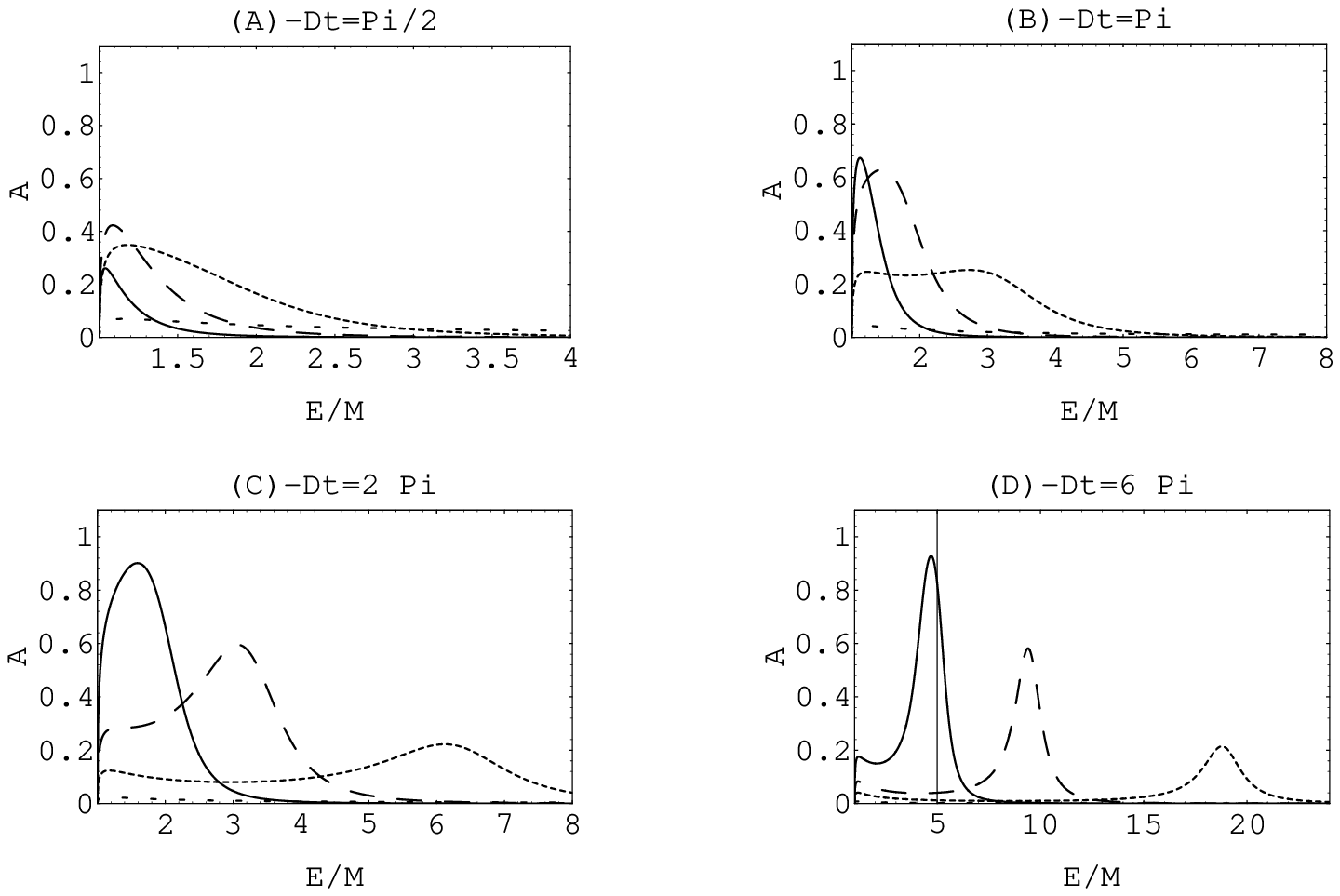,height=14cm}
\caption{The asymmetry A as a function of $E_l/m_0$. 
Continuos line: $\mu=2$, 
dashed lines: respectively 
$ \mu=1,1/2,1/10$. 
From A to D: $\Delta\theta=\pi/2, \pi, 
2\pi, 6\pi$. 
 }
\label{f1}
\end{figure}

     In Fig.(\ref{f1}) the asymmetry A is plotted as a function of the
     dimensionless quantity $E_l/m_0$ and some fixed values of $\Delta\theta$.
     For $\Delta\theta=\pi$ (plot B) the results obtained here coincide 
     with the      numerical result obtained in \cite{nel1} (Fig.  2).  
     For a realistic case  where   
   $\lambda\sim M_H\sim M_W$ and $m_0\sim m_{top}$, 
 the value of $\mu\equiv m_0\delta\equiv m_0/\lambda\approx 2-2.5$.   
 For this range of $\mu$ (continuous curves in Fig.(\ref{f1}))
    the asymmetry A is 
non-neglible for a large  $E/m_0$ range.
In general A is negligible for fermion 
masses corresponding to $\mu \lessapprox 10^{-1}$ which can be translated
to the condition
$m_0\lessapprox 2 m_{bottom}$ for  not so heavy Higgs masses.

     As the phase
     difference $\Delta \theta$ increases, the height of the peaks keeps
     roughly
     unaltered but their position moves:
     $$\left  (E_l/m_0\right )_{peak}=\mid \Delta\theta\mid/2\mu.$$ 

     For large differences 
     $E_l/m_0-\Delta\theta/2\mu$, $\overline{R}$ becomes 
     neglible and the 
     asymmetry $A\sim\mid R\mid^2$. $\overline{R}$ is
     non-neglible only for 
     $E/m_0$ and $\Delta\theta$ small enough, such that 
     the imaginary part of 
     the first parameter of the Hypergeometric function 
     in Eqs.(\ref{e6024}) approaches zero.

     For higher values of $\Delta\theta$ 
     some resonant effect becomes apparent, the peaks get sharper and the
     asymmetry is only non-negligible around them.  This is particularly
     evident in the last plot (D).  This effect could lead to an essentially
     monochromatic asymmetry in scenarios with highly oscillatory or random
     phase differences.
     For high $\Delta\theta$, the asymmetry quantity can be approximated by the
     expression 
     \begin{equation} 
     A\approx \delta \left(
     E_l/m_0-\Delta\theta/2\mu\right ) f\left (\mu,\Delta\theta \right)
     \label{e1036} 
     \end{equation} 
     where $\delta(x)$ is a delta-like peaked function and
     $f(\mu,\Delta\mu)$,      the value of A at peak, is
     easily computable from Eqs.(\ref{e1028}), 
     it is  of interest only its
     general behavior: 
\begin{equation} 
f(\mu,\Delta\theta)\to\left \{ 
\begin{array}{cc}
\mu^2/2\Delta\theta  &  \hbox{for}\; \mu\to 0 \\
                           1 &  \hbox{for} \; \mu\to \infty
\end{array}
\right .
\end{equation} 
   
\begin{figure}[p]
\centering\hspace{0.8cm}
{\epsfig{file=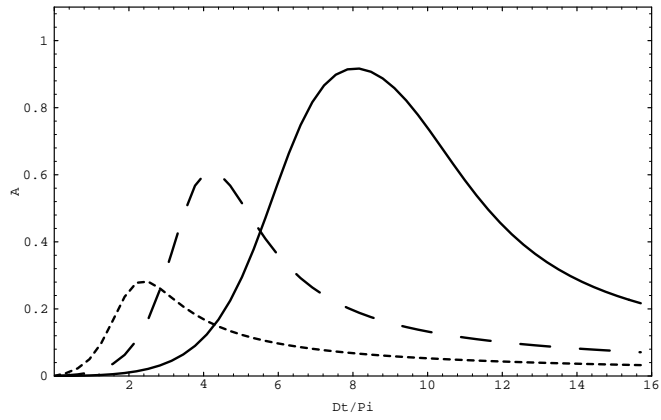,height=12cm}} 
\caption{Asymmetry $A$ as a function of $\Delta\theta$ and
different $\mu$ as before. For each curve, the energy is
 choosen as $E_l/m_0=2$.}  
\label{f2}
\end{figure}

     The resonant behavior is again evident in
     Fig.(\ref{f2}) where we plot A as a function of the 
     parameter $\Delta\theta$
      for various $\mu$      as before.  
     The peaks in the asymmetry function appear for values such that 
     $\Delta\theta_{peak}=2 E_l\delta$.

\section{Chiral charge flux through the bubble wall
and the Baryon asymmetry.}

The chiral charge flux in front of the wall in the symmetric phase and 
in the wall frame is
given by \cite{nel1}:

\begin{eqnarray}
F &=&\frac{1}{2\pi^2\gamma}\int_0^\infty dp_l\int_0^\infty p_t dp_t (f^s(-p_l,
p_t)-f^b(p_l,p_t)) A(m_0,\lambda,p_l)
\label{e1037} 
\end{eqnarray}

A is the asymptotic 
asymmetry defined in the previous section and 
the thermal equilibrium fermion flux densities $f^s,f^b$ 
in the symmetric and broken phases respectively
(neglecting Fermi blocking factors) are:

\begin{eqnarray}
  \label{e1}
  f^s&=&\frac{(p_l/E^s)}{\exp(\gamma (E^s-u p_l)/T)+1}, \; 
E^s=\sqrt{p_l^2+p_t^2} \nonumber \\
  f^b&=&\frac{(p_l/E^b)}{\exp(\gamma (E^b+u p_l)/T)+1}, \; 
E^b=\sqrt{p_l^2+p_t^2+m_0^2} \nonumber
\end{eqnarray}

The wall
velocity in the fluid frame is given by u ; $\gamma=1/\sqrt{1-u^2}$; $p_l,p_t$ 
are the longitudinal
and transverse momentum of the fermion with respect to the wall.
T is the equilibrium temperature.
The use of asymptotic expressions for $f^s$ and $A$ is questionable, the 
 position dependent mass implies the existence of a density gradient even
at thermal equilibrium. The Formula (\ref{e1037}) would be valid
at far  distances  from the wall and in absence of inter-particle collisions. We can expect that the formula is reasonable valid as long as the wall
width is small compared to the mean free path in the plasma, so effects inside the wall do not play an important role. 

Defining dimensionless momentum and wall width
 $p_l/m_0=x; p_t/m_0=y, \epsilon=T/\lambda$, defining also 
$\kappa= \mu \gamma/\epsilon$ and
integrating with respect $p_t$, we arrive to the expression:

\begin{equation}
  \label{e1042}
F=T^3 \frac{\mu^2}{2 \pi^2 \gamma^2\epsilon^2} F^\ast(\mu,\kappa,u)
\end{equation}
Where $F^\ast$ is the function defined by the integral:
\begin{equation}
  \label{e1043b}
F^\ast(\mu,\kappa,u)=\int_0^\infty dx\ x A(\mu,x)
\left (\log
\frac{1+\exp( -\kappa(1-u) x)}{1+\exp\left(-\kappa\sqrt{1+x^2}-\kappa u x\right)}
\right )
\end{equation}

The dependence with the CP violating phase $\Delta\theta$ is fully contained in the asymmetry $A(\mu,x)$.
The integrand in Eq.(\ref{e1043b}) is defined positive if $u>0$. For negative
$u$ (as it could happen in late wall contraction phases) its sign changes at a 
certain $x_0=1/(2\sqrt{u^2+\mid u\mid})$.
For large $x$ the integrand multiplying A in Eq.(\ref{e1043b}) 
 approaches 
zero exponentially for any value of the parameters,
 the most important contribution to the integral
comes from the values of $A(x)$ with
 $x$ relatively small.
This implies that in the case $F^\ast$ might become negative, its absolute
value would be relatively small.
It is expected that the resonance observed  in the previous section tends
to be washed out due to this behavior: the resonance in A appears at large x, just when the rest of the integrand is exponentially smaller.
The resonance in A could have a more significant effect in non-equilibrium
cases: for flux densities $f^s,f^b$ with large tails at large momentum.

\begin{figure}[p]
\centering\hspace{0.8cm}
{\epsfig{file=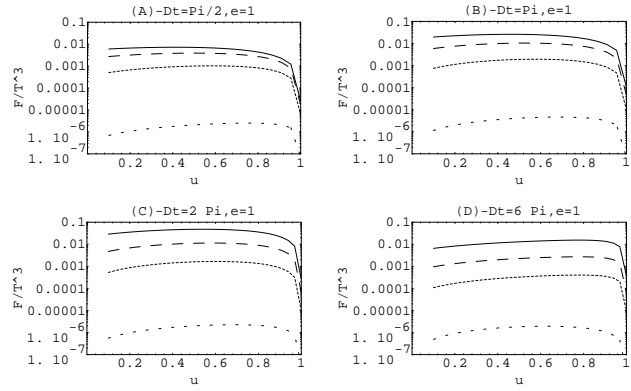,height=12cm}} 
\caption{ Chiral 
charge flux (Eq.(\protect\ref{e1042}))
as a function of the wall velocity u, for several $\Delta\theta$ and
$\mu$ (continuous and dashed lines as in previous figures). In all 
the figures 
the adimensional wall width 
$\epsilon=1$.}  
\label{f3}
\end{figure}

We plot in  
Fig.(\ref{f3}) 
 the chiral flux divided by the cubic temperature
given by the Eqs.(\ref{e1042}-\ref{e1043b}) as a function of u.
$F/T^3$ is rather independent of u and $\Delta\theta$
in  the logarithmic scale of the figure, except for $u\to 1$.
It is however 
strongly dependent in $\mu$, it changes a few orders of magnitude for
$\mu$ varying changing in the range $1-10^{-1}$.

For large $\Delta\theta$, using the formula (\ref{e1036}), one 
 obtains  the approximate expression for the integral $F^\ast$:
\begin{equation}
  \label{e1040}
F^\ast\approx f(\mu,\Delta\theta)
\log
\frac{1+\exp\left ( -\kappa(1-u) \sqrt{(\Delta\theta/2\mu)^2-1}\right )}
{1+\exp\left (-\kappa\left(\Delta\theta/2\mu+ u\sqrt{(\Delta\theta/2\mu)^2-1}\right)\right)}
\end{equation}

For a large value of $\Delta\theta$ but keeping the quantity $\kappa 
\Delta\theta /2 \mu\equiv \gamma \Delta\theta/2\epsilon$ 
relatively small, $F^\ast$, and so $F$, 
are  proportional to $ \Delta\theta$
\begin{equation}
\label{e1044}
F^\ast=
\frac{\kappa (1+u)}{2} f(\mu,\Delta\theta) \frac{\Delta\theta}{2\mu}
\end{equation}
This behavior 
coincides essentially with the proportionality in $\Delta\theta$ obtained in \cite{fun3}.

In the opposite regime, for a large quantity $\kappa \Delta\theta/2\mu$, 
$F^\ast$ approaches zero
exponentially:
\begin{equation}
\label{e1045}
\label{e1041}
F^\ast\approx
2f(\mu,\Delta\theta) e^{-\kappa \Delta\theta/2\mu} \sinh \left (\kappa u \Delta\theta/2\mu\right )
\end{equation}

\begin{figure}[p]
\centering\hspace{0.8cm}
{\epsfig{file=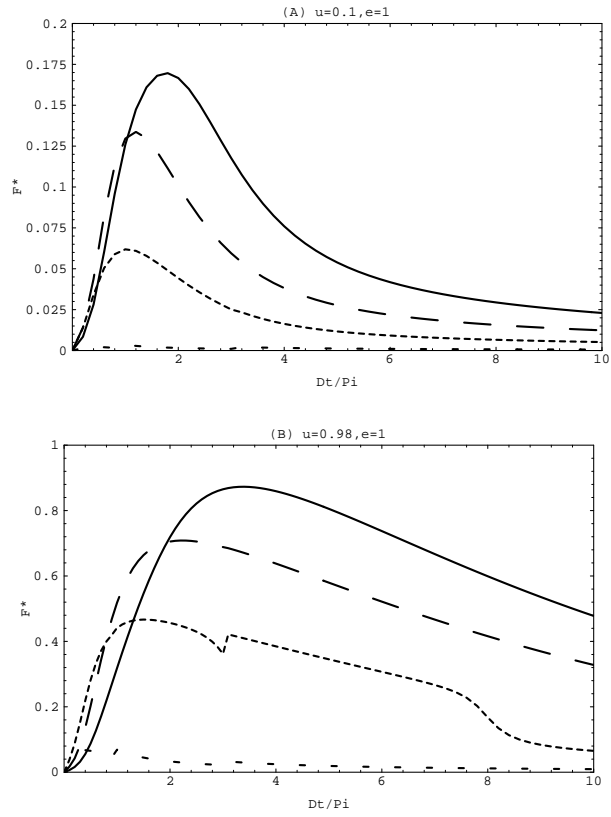,height=12cm}} 
\caption{Dependence with $\Delta\theta$ for different 
$\mu$ ratios.}
\label{f6}
\end{figure}

We plot $F^\ast$ as a function of $\Delta\theta$ in
 Figs.(\ref{f6}) (given by the the exact 
Formula (\ref{e1043b}). The small and large $\Delta\theta$ behavior observed
in the figure are clearly given by the Eqs.(\ref{e1043}-\ref{e1044}) respectively.
It is interesting to compare Fig.(\ref{f6}) with  Fig.(\ref{f2}). In agreement with what was expected from the form of the integral (\ref{e1043b}), the resonances are lost here. However, It remains
still a non obvious remarkable behavior. 
$F^\ast$ (and then $\rho_b/s$ to be defined later) is 
$\propto\Delta\theta$ in a surprisingly sizeable range.
after reaching 
 a maximum for $\Delta\theta$ around
 $2-4 \pi$ for practically any value of the rest of the parameters, $F^\ast$ goes quickly to zero for higher
values.

\subsubsection{The baryon asymmetry}

The baryon asymmetry/entropy ratio of the Universe through 
the sphaleron process
is given in the charge transport scenario with two Higgs doublets
by \cite{nel1}: 
\begin{equation}
\frac{\rho_B}{s}=\frac{4}{5}\frac{\Gamma_B}{T}\frac{\tau}{u}\frac{F}{s}
\label{e1043}
\end{equation}
Where
\begin{itemize}
\item
$\Gamma_B$ is the sphaleron transition rate per unit of volume
in the symmetric phase  \cite{nel4}:
\begin{equation}
\Gamma_B\simeq 3 \eta \alpha_w^4 T^4
\end{equation}
with an unknown numerical factor $\eta\approx 0.1-1$ \cite{amb1}. 
Note however 
that a violation rate $\sim O(\alpha_w^5 T^4)$
 has been proposed recently \cite{arn1}.
The introduction of an additional factor 
$\alpha_w$ would not change
greatly the final results, or at least its order of magnitude,
but it is an indication of the global incertitude still existing
 on $\Gamma_B$.

\item
$\tau$ is a typical transport time within which the scattered fermions are
captured by the wall and in which they can be converted into baryon number
\cite{nel1}  $ \tau\simeq l/u$
 with l the mean free path of the corresponding particle.
In some works $\tau$ is computed numerically \cite{nel1}.  We will 
parametrize it in the form
$$l=\frac{D}{T}$$
 D can be interpreted as a   diffusion-like constant 
with a magnitude in the order $D\approx 1-20$
 (for example $D\approx 4-6$ in \cite{bon1,nel4}).

\item
The entropy density is given by $s=2\pi^2 g_\ast T^3/45 $
with $g_\ast\approx 100$, the effective degrees of freedom of the relativistic
particles at the electroweak phase transition.
\item F is the charge flux given by Eq.(\ref{e1042}) which contains all the
dependence on $\Delta\theta$ through the integral $F^\ast$.
\end{itemize}

Inserting all the factors
 the baryon/entropy ratio is given explicitly by:
\begin{equation}
\frac{\rho_B}{s}\approx 5.10^{-9}\times (\eta D)\times\frac{\mu^2 F^\ast}{u^2 
\epsilon^2\gamma^2}
\label{e1049}
\end{equation}
$F^\ast$ is a function only of the adimensional parameters $\epsilon,\mu,u$ and
$\Delta\theta$. The 
explicit dependence with the temperature T has vanished in this formulation. 
The range of validity of Eq.(\ref{e1043})
is limited by the simplicity of charge transport model. It should be valid
for $ \Gamma_B\tau/T^3<<u $ (\cite{nel1}) or
$\eta D<< 10^6 u^2$. With $\eta\approx 1, D\approx 10$, we would have 
then the condition $u>>10^{-2}$.
In the other extreme
the formula lack a priori validity for $u\to 1$. In this case there is 
not time for the fermion in front of the wall to thermalize and the 
flux densities defined in the previously can not be used. 
For those cases  a detailed Boltzman transport treatment should be performed instead. Note however that, according to previous arguments taking into account
the dependence of A at large momentum, the thermal case would act as
 a lower limit.
Another constraint comes from the
condition for 
the validity of formula (\ref{e1037}): the wall width should be smaller
that the mean free path of the fermion or
 $\epsilon=T/\lambda\lessapprox D$.

\begin{figure}[p]
\centering\hspace{0.8cm}
{\epsfig{file=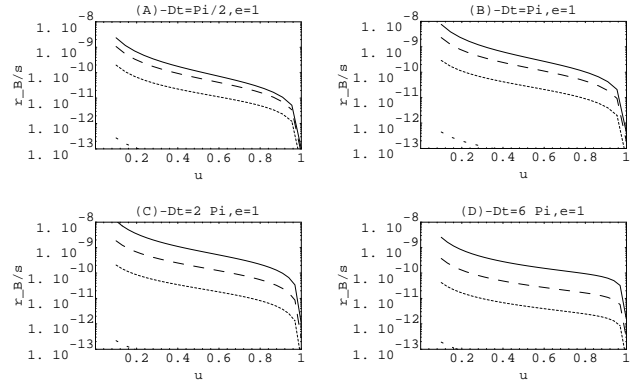,height=12cm}} 
\caption{$\rho_B/s$ ratio (Eq.(\protect\ref{e1049}), $\eta D=1$) 
as a function 
of the wall velocity u (see explanation for the previous figure).}  
\label{f4}
\end{figure}

The matter content of the observed universe
is given by $\rho_B/s\simeq 10^{-10}$
(from nucleosynthesis calculations, see \cite{nel1} and references
therein).
Figure (\ref{f4}) shows the ratio $\rho_B/s$. 
For large $u$, $\rho_B/s$ follows the behavior of F: decreases and approaches
 zero for $u\to 1$.
For small u, as long as the equation is still valid, the divergent behavior
$\sim1/u^2$ is apparent.
According to Fig.(\ref{f4}) a value $\rho_b/s\approx 10^{-10}$
 may be explained by this
mechanism for weak scale baryogenesis for a 
wide range of parameters, including a large variation in the rather
unknown product $\eta D$, as long as $\mu\sim 0.5-2$. 
$\mu$ is the main parameter needed to be tuned.

\begin{figure}[p]
\centering\hspace{0.8cm}
{\epsfig{file=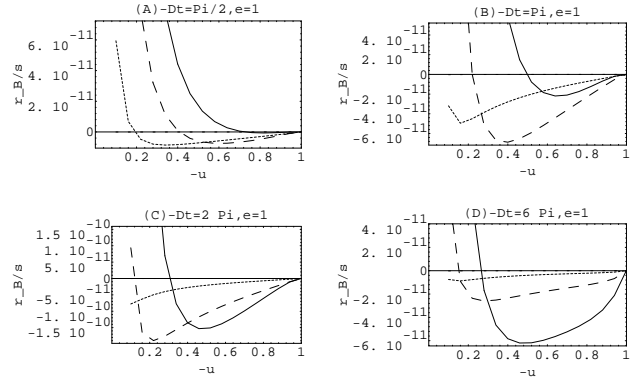,height=12cm}} 
\caption{$\rho_B/s$ ratio , (Eq.(\protect\ref{e1049}), $\eta D=1$) 
 for negative wall speeds (contracting walls).}
\label{f4b}
\end{figure}

 In Figure (\ref{f4b}) the same $\rho_b/s$ is plotted as a function of negative
wall speed (contracting wall case). 
In agreement with previous considerations $\rho_B/s$ is 
of opposite sign in this case and  relatively smaller in absolute value.
The effect is more important for $\mu\approx 1-2$, $\Delta\theta$ large 
and medium speeds. 
Note that the pattern of the $\mu$ dependence is different here than that one
appearing in Fig.(\ref{f4}).
For a detailed knowledge of the final $\rho_B/s$ an integration over the full history of the bubble may be neccesary.

\begin{figure}[p]
\centering\hspace{0.8cm}
{\epsfig{file=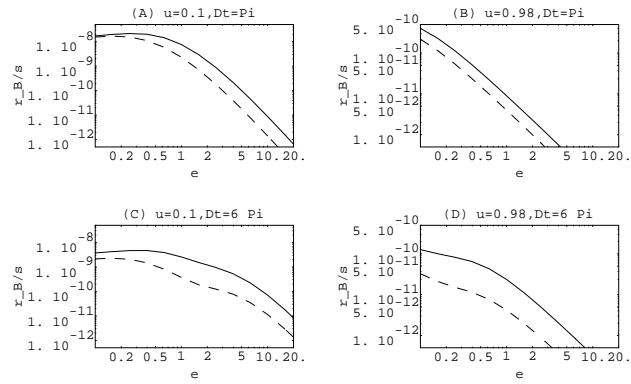,height=12cm}} 
\caption{Dependence with the thickness of the wall $\epsilon=T/\lambda$
. For high and low values of the wall velocity and $\Delta\theta$. Continuos line: $\mu=2$, dashed line $\mu=1$.}  
\label{f5}
\end{figure}

The dependence with the adimensional 
thickness of the wall ($\epsilon=T/\lambda$) 
is shown in Fig.(\ref{f5}) 
( cf. with 
Fig.(4) in \cite{nel1}).
$\epsilon$  corresponds approximately to the inverse Higgs mass scale 
and should be independent of T ($\lambda\sim T$)  in most models.  
The dependence is rather modest for thin walls ($\epsilon<1$). For relatively thicker walls $\epsilon>5-10$ the suppression is very important: values as
$\rho_B/s<<10^{-11}$ are reached in this region.

The dependence with the CP violating phase $\Delta\theta$ of $\rho_b/s$ is
contained totally in $F^\ast$ and appears 
in  Fig.(\ref{f6}). Although interesting qualitatively, 
it is clear from a numerical point of view
that the maximum variation , a factor 4 from $\Delta\theta\approx\pi/2$ to $\Delta\theta\approx 2\pi$, is still small in comparison with all the other
incertitudes appearing in the present knowledge, observational and theoretical,
of $\rho_b/s$. An important conclusion from Fig.(\ref{f6}) is that,
 although decreasing, 
$\rho_b/s$ still keeps the right order of magnitude
even for very large $\Delta\theta$.

\section{Summary and conclusions}

As conclusion,  we were able to solve the Dirac equation  with a space
dependent complex mass term which, although not a solution
of the equations of motion,  reproduces  
the expected variation in module and in complex phase  
of the average Higgs
field across the wall.
From the solution of the Dirac equation the particle-antiparticle 
transmission asymmetry  is computed.
The analytical results presented here confirm previous numerical 
computations for small $\Delta\theta$ and 
predict an unexpected behavior for highly oscillatory phase
fields.

We have computed the chiral flux and the baryon asymmetry.
For a large set of parameters this is compatible with the ''experimental''
value $\rho_B/s\sim 10^{-10}$. This is true even for extremely large  
values of CP violation $\Delta\theta$.
The resulting baryon asymmetry for contracting walls has been estimated for the
first time to our knowledge.

The main uncertainties come from the rate of anomalous baryon number violation at high temperature and from the method itself.
In view of the dependence on $\epsilon$ and $\Delta\theta$, It is clear that the plausibility of this model depends only on the existence of a massive fermion (as the Top quark) and is rather independent
of the values of the Higgs mass and the critical temperature of the weak transition as long as they keep in a reasonable range.

A more detailed microscopic treatment based in the Boltzman equation is clearly
advisable. The solution of the Dirac equation obtained here is particularly
suitable for this purpose. Although the mass term which has been used is 
somehow artificial,  is a suitable toy model to 
study large CP violation environments. 
This solution can be used  as a starting point to obtain
numerical correction for more realistic mass fluctuations.

The  method  used here can be used with little modifications  to solve
the Dirac equation for a purely local dependent phase mass term ($\lambda\to 0,
\Delta\theta\lambda\not=0$ in Eq.(\ref{e1103}) or equivalently $\sigma$ purely 
imaginary in Eq.(\ref{e1110})). 
In order to circumvent convergence problems this must be done
  taking the
limit $\Re \lambda\to 0$ in the finite propagation time Eq.(\ref{e1125}).

  \appendix

  \section{Appendix: some old and new formulas about Hypergeometric Functions}

  The generalized Hypergeometric function \cite{grad} is defined by
  
\begin{equation}
  {}_p F_q (a_1,\dots,a_p,b_1,\dots,b_q; z)=\sum_{n=0}^\infty 
  \frac{(a_1)_{(n)}\dots (a_p)_{(n)}}{(b_1)_{(n)}\dots (b_q)_{(n)} }
    \frac{z^n}{n!}
  \label{a140}
  \end{equation}

where    the Pochammer symbol is 
\begin{equation}
  (z)_{(n)}=\Gamma(n+z)/\Gamma(z)
  \label{a110}
  \end{equation}

in particular
\begin{equation}
  \mbox{${{}_0 F_1}$}(b;z)=\sum_{n=0}^\infty \frac{1}{(b)_{(n)}} \frac{z^n}{n!}
  \label{a100}
  \end{equation}

The derivative of ${}_0F_1$ is again a ${}_0 F_1$ function:
\begin{eqnarray}
    \frac{d\ \mbox{${{}_0 F_1}$}(\gamma;z)}{dz}&=&\frac{1}{\gamma}\
  \mbox{${{}_0 F_1}$}(1+\gamma;z)  \label{a120} 
\label{a160}
\end{eqnarray}

The function ${}_0 F_1$ is related to the Bessel Functions
by the formula
\begin{equation}
J_n(z)=\frac{(z/2)^n}{\Gamma(1+n)} {}_0F_1\left (1+n; -\frac{z^2}{4}\right )
\end{equation}

For a $Q$ as 
given by Eq.(\ref{e6009}) is traceless, $det \ Q(z,z_0)=det\ Q(0,0)=1$, or
$\mid F\mid^2-\mid G\mid^2=1 $.
The formulas (\ref{e6024}) are valid for any
$E,\Delta\theta$ real, $k$ complex, we obtain:
\begin{equation}
\mid {}_0 F_1\left (\frac{1}{2}+\epsilon i;  x^2\right )\mid^2-\frac{x^2}{\mid
1/2+\epsilon i\mid^2} \mid {}_0 F_1\left (\frac{3}{2}+\epsilon i; x^2\right 
)\mid^2=1
\end{equation}
to be compared with the similar formulas obtained in \cite{emi1} for the
absolute values of the generalized hypergeometric functions ${}_n F_n$.

In fact, with little changes, one could compute U for any general matrix V
with diagonal terms equal to zero and non-diagonal terms of general exponential
type not neccesarily equal.
In the particular case of V hermitic: $V_{12}=V_{21}^\ast=k$,  $U_{red}$
 is unitary and of the form
\begin{eqnarray}
U_{red}(z_0)&=&\pmatrix{
 \overline{F} & \overline{G} \cr -\overline{G}^\ast & \overline{F}^\ast \cr }
\label{e6027}
\end{eqnarray}
with $\overline{F},\overline{G}$ given by Eq.(\ref{e2122}).

By the unitarity of the matrix (\ref{e6027}), 
$\mid F\mid^2+\mid G\mid^2=1 $. And we get the formula:
\begin{equation}
\mid J_{-\frac{1}{2}+\epsilon i}(x)\mid^2+\mid J_{\frac{1}{2}+\epsilon
i}(x)\mid^2=\frac{2\cosh \pi \epsilon}{\pi x}
\end{equation}
For $\epsilon=0$ this formula reduces to the trigonometric
 case involving the well known 
Bessel functions of order $1/2$: $J_{-1/2}(x)=\sqrt{2/(\pi x)}\cos(x); \
J_{1/2}(x)=\sqrt{2/(\pi x)} \sin(x) $.

\vspace{1cm}
{\Large{\bf Acknowledgments.}}

The author would like to thank to Peter Minkowski for many enlightening discussions. This 
work has been supported in part by the Wolffman-Nageli Foundation (Switzerland)
and by the MEC-CYCIT (Spain).

\newpage

\end{document}